\documentclass{article}
  
\usepackage[utf8]{inputenc}
\usepackage[T1]{fontenc}
\usepackage[mathscr]{euscript}
\usepackage{geometry}
\usepackage{amssymb}
\usepackage{amsmath}
\title{A note on singular and non-singular black holes}
\author{Stefano Chinaglia\footnote{e-mail: s.chinaglia@unitn.it}  and Sergio Zerbini\footnote{e-mail:sergio.zerbini@unitn.it} \\
\smallskip \\
\it Dipartimento di Fisica, Università di Trento \\
\it Via Sommarive 14, 38123 Trento, Italia \\
\smallskip \\
\it TIFPA (INFN) \\
\it Via Sommarive 14, 38123 Trento, Italia}
\date{\today}

\begin{document}
\maketitle
\begin{abstract}

An attempt is made in order to clarify the so called regular black holes issue. It is revisited that if one works within General Relativity minimally coupled with non linear source, mainly of electromagnetic origin, and within a static spherically symmetric ansatz for the metric, there is still room for singular contribution to the black hole solution. A reconstruction method is proposed and several examples are discussed, including new ones. A possible way to obtain a non singular black hole is 
introduced, and in this case,  several known examples are re-discussed, and new ones are provided.

\end{abstract}

\section{Introduction}

Black holes are solutions of Einstein equations of General Relativity (GR) and, as it is well known, their existence  have been recently confirmed by the first detection of associated gravitational waves  by Ligo-Virgo collaboration \cite{A}. Thus, they exist in our universe. On the other hand, GR, under reasonable hypothesis on stress tensor, predicts  a central singularity inside them at $r=0$. This is a well known fact.

 Nature is indeed expected to be singularity-free. Thus, the presence of singularities in GR  is usually regarded as a breakdown of the classical theory in some regime, and modifications of the theory 
are thought  to be required, and these modifications are likely to be described by quantum effects. 
In fact, within  FRWL cosmologies, quantum loop cosmology leads to a resolution of the initial Big Bang GR singularities \cite{bojo}.

On the other hand, one may try to work with an effective theory approach. This is the idea that lead Born and Infeld to propose a non-linear electrodynamics  (NED) \cite{Born_Infeld}, a generalization of  Maxwell theory.  There they were able to resolve the singularity of  the Coulomb potential for $r \rightarrow 0$. One might expect that something similar could happen in GR.  

We recall that a sufficient condition to deal with a non singular black hole solution is the requirement that there exists for small $r$, at least,  a de Sitter(AdS)  core. This is called   Sakharov criterion  \cite{Sakharov},  namely, for small $r$, the Einstein tensor $G_{\mu \nu} \simeq \Lambda_0 g_{\mu \nu}$, with $\Lambda_0 $ not vanishing.

As stressed by several authors, see for example \cite{abby,Hayward_1} the absence of central singularity renders also the semi-classical Hawking radiation and related information loss issues less problematic. As a consequence, it is of certain interest to investigate the non singularity issue in gravitation theories.

The first author to propose a non singular  black hole solution was Bardeen \cite{Bardeen}. 
Then several papers followed. A non exhaustive list of  papers is \cite{Bronnikov} \cite{Elizalde} \cite{Dymnikova} \cite{Hayward_1} \cite{NSS} \cite{ANSS} \cite{Modesto}\cite{Dymnikova_2} \cite{Culetu} 
\cite{Maeda} \cite{Pradhan} \cite{Ma} \cite{Johannsen} \cite{Rodrigues}\cite{Fan}\cite{Frolov}. For a  review, see \cite{ans}.

In discussing the non singularity issue, several  approaches have been used. For example, one  may  modify the geometric sector of the Einstein Equations, or the matter content.

Among these, \cite{Elizalde} \cite{NSS} \cite{ANSS} \cite{Modesto} \cite{Maeda} work on the gravitational sector. \cite{NSS} \cite{ANSS} smear the singularity introducing non commuting coordinates, i.e. $[x_{\mu}, x_{\nu}] \neq \delta_{\mu \nu}$, while \cite{Maeda}  directly modifies the Einstein tensor. 

 On the other hand, \cite{Bronnikov} \cite{Dymnikova} \cite{Dymnikova_2} \cite{Culetu} \cite{Pradhan} follow the second approach,  mainly working within  NED.

In the late '90 E. Ayon-Beato and A. Garcia proposed a scheme to generate (even regular) solutions from a NED Lagrangian minimally coupled to the standard Lagrangian of GR \cite{Beato}. They successfully applied their scheme to build up a modified Reissner-Nordström, Bardeen-like, solution, later identified as a magnetic mono-pole solution \cite{Beato_1}. Their scheme was later generalized by I. Dymnikova \cite{Dymnikova} for spherical and static solutions. Further properties can be found in
\cite{NBS}.

The main aim of NED is to look for some non-linear modifications to standard electrodynamics, able to remove the central singularity of the usual Reissner-Nordström  (RN) black hole. Unfortunately, this is very difficult to be implemented in practice and very rarely, at Lagrangian level, one gets full analytic results. For this reason, the so called dual P approach \cite{GSP} has been proposed, in which central role is not played by the NED Lagrangian.   

We should also  mention the attempts based on the introduction of  non minimal coupling between the gravitational and the matter Lagrangian:  Dereli and Sert \cite{Dereli}, Sert \cite{Sert}, Balakin and Lemos \cite{Balakin}, Balakin and Zayats \cite{Zayats} and, recently, Balakin, Lemos and Zayats \cite{BLZ}. \cite{Dereli} and \cite{Sert} write the Lagrangian in the form $\mathscr{L} = R - Y(R)I$, where $Y(R)$ is a given function of $R$, while \cite{Balakin}, \cite{Zayats} and \cite{BLZ} follow the lines of Horndeski \cite{Horndeski} and of Drummond and Hathrell \cite{Drummond} and use $\mathscr{L} = R + F^{(a)}_{\mu\nu} F^{(a)\mu\nu} + \mathscr{R}^{\mu\nu\rho\sigma} F^{(a)}_{\mu\nu} F^{(a)}_{\rho\sigma}$, where $\mathscr{R}^{\mu\nu\rho\sigma}$ is a complicated object worked out from the Riemann tensor. While the approach of \cite{Dereli} and \cite{Sert} is not able to produce regular black holes, the one  of \cite{BLZ} actually does.

The aim of present paper is to revisit and discuss again the presence of central singularity for  static spherically symmetric space-times. We stress that we do not just perform a purely formal discussion, but we work within the framework of a specific Lagrangian approach, and the properties of related stress-energy tensor such Weak Energy Condition (WEC) may be discussed.

The paper is organized as follows.  In  Section 2, first we review the singularity issue in the so called fluid effective approach, namely Einstein gravity plus some unspecified effective source. In Section 3, we present the NED scheme of Beato-Garcia-Dymnikova (BGD), discussing some examples. In sec. 4 we provide 
a reconstruction scheme for NED approach. In Section 5 we present an effective approach to some classes of non singular black holes.  The paper ends with the conclusions in Section 6.

\section{The effective approach:  equations of motion}

We start considering the  static spherically  symmetric metric of the form

\begin{equation}
\label{metric}
ds^2 = -f(r,t)dt^2 + \frac{1}{g(r,t)} dr^2 + r^2 d\Omega^2_k\,,
\end{equation}

where $f(r,t)$ and $g(r,t)$ are some arbitrary functions of $r$ and $t$ and $d\Omega^2_k$ is the volume of a 2-sphere ($k=1$) or a hyperbolic manifold ($k=-1$) or a torus ($k=0$). Although topology does not play any role in our argument, we recall that topological black holes have been widely studied since the works of Brill, Luoko and Peldàn \cite{BLP}, Mann \cite{Mann} and Vanzo \cite{Vanzo} and there is no problem to generalize the discussion of the spherical case to different topologies, and  the dependence on time may be absorbed in a redefinition, see for example \cite{chinaglia16}.

Furthermore, we may simplify the discussion taking  $f(r)=g(r)$ and $k=1$, and  the metric ansatz reads

\begin{equation}
\label{metric1}
ds^2 = -f(r)dt^2 + \frac{1}{f(r)} dr^2 + r^2 dS^2\,.
\end{equation}

In order to treat possible extented gravitational models, we assume an effective  model description and related generalized form of gravitational equations with $ G=1$

\begin{equation}
\label{ege}
G_{\mu \nu} = -8\pi \tau_{\mu \nu}\,,
\end{equation}
where $\tau_{\mu \nu}$ is an effective stress tensor.

We remind that a sufficient condition to deal with some non singular black hole is that there exists  a regular center, namely, for $r \rightarrow 0$, $f(r) = 1 + a r^n$, with $n \geq 2$, Sakharov criterion being equivalent to $n=2$.

Let us  introduce the effective density $\tau^0_0=\rho$, the effective radial pressure $p_r=\tau_1^1$ and the effective tangential pressure $p_T=\tau^2_2=\tau^3_3$. 
 As a result, the effective gravitational equations reduce to
\begin{equation}
\label{ege1}
rf'+f-1 = - 8\pi r^2 \rho\,,
\end{equation}
\begin{equation}
\label{ege2}
rf'+f-1 = 8\pi  r^2 p_r\,,
\end{equation}
\begin{equation}
\label{ege3}
rf''+2f = 16 \pi r p_T\,.
\end{equation}
As a consequence,  one may consider only equation (\ref{ege1}), since one has
\begin{equation}
\label{ege4}
p_r = - \rho\,, \quad P_T=-\rho-\frac{r}{2}\rho'\,,
\end{equation}
 these being equivalent to stress tensor conservation $\nabla^{\mu} \tau_{\mu \nu}=0$ (Bianchi identity).

We may  rewrite (\ref{ege1})  as 
\begin{equation}
\label{ege5}
rf'+f=1  - 8\pi r^2 \rho\,.
\end{equation}

First let us suppose that the effective density $\rho$ does not depend on $f$ and $f'$. Thus, one has to deal with an inhomogeneous first order linear differential equation. The general solution is
\begin{equation}
\label{ege5}
f(r) =f_0(r)+f_1(r)\,,
\end{equation}
where $f_0(r)$ is the general solution of homogeneous equation $rf_0'+f_0=0$, namely
\begin{equation}
\label{ege6}
f_0(r)=-\frac{C}{r}\,,
\end{equation}
in which $C$ is an {\bf arbitrary} constant of integration, while $f_1(r)$ is a particular solution of inhomogeneous differential equation which we may take in the form
\begin{equation}
\label{ege7}
f_1(r)=1-\frac{8\pi}{r}\int_0^r dr_1 r_1^2\rho(r_1)\,.
\end{equation}
As a result, the general solution reads
\begin{equation}
\label{ege8}
f(r)=1-\frac{C}{r}-\frac{8\pi}{r}\int_0^r dr_1 r_1^2\rho(r_1)= 1-\frac{C}{r}-\frac{2m(r)}{r} \,.
\end{equation}
The quantity $m(r)$ is called mass function. 

If the effective density $\rho$ does not depend on $f(r)$, and its derivatives, it cannot depend on the constant $C$, and as a consequence the model described by above gravitational equation contains the $\frac{C}{r}$ term, and this leads to a central singularity in $r=0$, as soon as $C$ is not vanishing. This is  independent on the  contribution coming from $f_1(r)$. This result is known within the NED approach (see for example \cite{breton} and original references cited therein).   

Furthermore,  it is not difficult to find conditions  on $\rho$ in order to have for $f_1(r)$ a de Sitter core, namely one has the sufficient condition

\begin{equation}
 \label{S}
\lim_{r \rightarrow 0} \rho(r)=A \,,
\end{equation}
which is obviously equivalent to
\begin{equation}
 \label{S1}
 r^2 \rho(r)=Ar^2+O(r^3) \,,
\end{equation}
for small $r$. Of course, the simplest choice is $\rho=\frac{3 H_0^2}{8\pi} $, $H_0$ constant and one has
\begin{equation}
 \label{ege9}
f_1(r)=1-H_0^2 r^2 \,,
\end{equation}
the general solution being the Kottler black hole solution, with the constant $C$ proportional to the black hole mass. 

Others well known examples are the choices $\rho(r)\equiv \rho_0 e^{-\frac{r^3}{a^3}}$, and $\rho(r) \equiv\frac{A}{(4\pi \theta)^{3/2}} e^{-\frac{r^2}{4 \theta}}$, which correspond to the Dynmikova \cite{Dymnikova92} and Spallucci et al. \cite{NSS} black holes.  All the black hole solutions associated with non linear electrodynamics belongs to this class \cite{Beato}. We shall discuss them in the next Section.

Within this framework, typically realized in Einstein gravity with minimally coupled matter, we may discuss the energy or mass associated with these solutions. There are many definitions of mass in this framework.  Here we shall make use of the quasi-local energy approach. We recall  that in the spherically static symmetric space-time  (SSS) case we are dealing with,   the so called Misner-Sharp energy is defined by (see for example  \cite{sean} and references therein)

\begin{equation}
 \label{ege10}
E_{MS}(r)=\frac{r}{2}\left(1-f(r)  \right) \,.
\end{equation}
If one has a SSS black hole, than there exists a $r_H>0$ such that $f(r_H)=0$. As a result, the black hole  mass may be identified by 
  
\begin{equation}
 \label{ege11}
M=E_{MS}(r_H)=\frac{r_H}{2} \,.
\end{equation}
In our case, the constant of integration $C$ is determined by  $f(r_H)=0$, namely  
\begin{equation}
 \label{ege12}
C=r_H- 2m(r_H)\,, \quad m(r_H)= 4\pi \int_0^{r_H} dr_1 r_1^2\rho(r_1)\ \,.
\end{equation}
As a result, the total black hole mass is
\begin{equation}
 \label{ege13}
M=\frac{C}{2}+ 4 \pi \int_0^{r_H} dr_1 r_1^2\rho(r_1)\,.
\end{equation}
Thus, in general there are present two contributions, the gravitational and matter or radiation ones. One has non singular black holes when $C=0$, and only the non gravitational mass is present. 

The use of MS mass is justified by the following considerations \cite{sean}. When a BH solution is present, we can compute the associate surface gravity, given by

\begin{equation}
 \label{ege14}
\kappa_H=\frac{f'_H}{2}\,.
\end{equation}
On the horizon Einstein equation gives

\begin{equation}
 \label{ege144}
\kappa_H=\frac{1}{2 r_H}- 4\pi \rho_H r_H\,.
\end{equation}

Recalling that in GR the horizon,  area and volume are given by $A_H=4 \pi r_H^2$ and $V_H=\frac{4}{3}\pi r_H^3$, one has 
\begin{equation}
 \label{ege15}
\frac{\kappa_H}{2\pi } d\left(\frac{A_H}{4}   \right)=d(\frac{r_H}{2}) - \rho_H dV_H    \,.
\end{equation}
We can interpret $\frac{\kappa_H}{2\pi }=T_H$ as Hawking temperature, and this is a well know robust result,    $S_H=\frac{A_H}{4}$ as Bekenstein-Hawking entropy, another well known robust result. With the identification of  $M=\frac{r_H}{2}$ as mass of the black hole and 
$ \rho_H dV_H $ work term according to  Hayward \cite{sean}, one gets  the First Law of black hole, namely 
\begin{equation}
 \label{ege16}
dM=T_H dS_H+\rho_H dV_H    \,.
\end{equation}

Finally, before concluding this section, few words about  WEC. We recall that  the WEC is satisfied if and only if \cite{Hawking_Ellis}

\begin{align}
\label{WEC}
& \rho \geq 0 \\
& \rho + p_k \geq 0 \ \ \ \ \ k=1,2,3\,,
\end{align}

where $\rho = T_t^t$ is the effective energy density and $p_k = - T_k^k$ the principal pressures. The issue if the WEC is satisfied or not by regular and NED regular solutions has been widely discussed, among the others, by I. Dymnikova in \cite{Dymnikova_WEC_1} and \cite{Dymnikova_WEC_2}. In particular, \cite{Dymnikova_WEC_1} finds some conditions a Lagrangian must satisfy, in order to fulfill the WEC. Later on, in the next section and in section 4, we will show that, under suitable conditions, all the solutions we found actually fulfill the WEC.

We also recall another physical constraint, namely the Dominant Energy Condition (DEC) \cite{Hawking_Ellis}, namely

\begin{align}
\label{DEC}
& \rho \geq 0 \\
& \rho + p_k \geq 0\,,  \quad  \rho-p_k \geq 0 \,, \ \ \ \ \ k=1,2,3\,,
\end{align}
which includes WEC.

\section{General Relativity and Non Linear Electrodynamics}
In this Section, we revisit a Lagrangian formulation which realizes the first case of the effective fluid approach previously discussed, the so called NED approach. This approach  has been  discussed in several papers, see, for example 
\cite{GSP,Beato,Dymnikova,breton,NBS}. 

The NED gravitational model is based on the following action 
\begin{equation}
\label{action}
\mathscr{I} = \int d^4 x \ \sqrt{-g} (\frac{R}{2} - 2\Lambda -\mathscr{L}(I))\,,
\end{equation}
where $R$ is the Ricci scalar, $\Lambda$ is a cosmological constant, and  $I=\frac{1}{4}F^{\mu\nu} F_{\mu\nu}$ is an electromagnetic-like tensor and $\mathscr{L}(I)$ is a suitable function of it. Recall that $F_{\mu\nu}=\partial_\mu A_\nu - \partial_\nu A_\mu$. We will only deal with gauge invariant quantities, and  we put $\Lambda=0$, because its contribution can be easily  restored. The equations of motion read
\begin{equation}
\label{eq1}
G^\nu_\mu  = -F_{\alpha\nu} \partial \mathscr{L} F^{\mu\alpha} + \mathscr{L}\delta^\nu_\mu 
\end{equation}
 \begin{equation}
\label{eq2}
\nabla^\mu (F_{\mu\nu} \partial_I \mathscr{L}) = 0\,.
\end{equation}
Another equivalent approach is called dual P approach and it is based on   two  new  gauge invariant quantities \cite{GSP}
\begin{equation}
\label{P}
P_{\mu\nu}\equiv F_{\mu\nu} (\partial_I \mathscr{L}(I))\,\,\,, P \equiv \frac{1}{4} P_{\mu\nu}P^{\mu\nu}\,,
\end{equation}
and 
\begin{equation}
\label{H}
 \mathscr{H} \equiv 2I(\partial_I \mathscr{L}(I)) - \mathscr{L}(I) \,,
\end{equation}
\begin{equation}
\label{eq_motion_2P}
\nabla^\mu P_{\mu\nu} = 0\,.
\end{equation}
In the following, we shall make use only of the traditional approach based on equations (\ref{eq1}) and (\ref{eq2}). 

Within the static spherically symmetric ansatz (\ref{metric1}) and from  (\ref{eq2}), one has 

\begin{equation}
\label{4}
\partial_r\left(r^2 \partial_I \mathscr{L}F^{0r}\right)   = 0\,,
\end{equation}
Since $I=\frac{1}{2}F_{0r}F^{0r}=- \frac{1}{2}F_{0r}^2$, one gets
\begin{equation}
\label{r111}
r^2 \partial_I \mathscr{L}   = \frac{Q}{\sqrt{-2 I}}\,,
\end{equation}
$Q$ being a constant of integration. As a result, within this  NED approach,  one may solve the generalized Maxwell equation. We shall make use of this equation and the the (t,t) component of the Einstein equation, which reads
\begin{equation}
\label{r}
G^t_t  =\frac{rf'+f-1}{r^2} = 8\pi\left( -2 I\partial_I \mathscr{L}+ \mathscr{L}\right)=-8\pi \rho\,.
\end{equation}
Introducing the  more convenient quantity  $X$
\begin{equation}
\label{x   }
X=Q\sqrt{-2  I}\,,          
\end{equation}
one may rewrite equation (\ref{r111}) as
\begin{equation}
\label{r2}
r^2 \partial_X \mathscr{L}   = 1\,.
\end{equation}
Furthermore, we have
\begin{equation}
\label{r4}
 \rho= X\partial_X \mathscr{L} - \mathscr{L}=\frac{X}{r^2}-\mathscr{L}  \,.
\end{equation}

Thus, when  $ \mathscr{L}(X) $ is given, then making use of (\ref{r111}), one may obtain $\rho=\rho(r)$, 
and gets the solution according  the general discussion on the previous Section.

\subsection{A class  of NED models }
As an example of this direct  approach, let us investigate the following class of NED models

\begin{equation}
\label{ln}
 \mathscr{L}(X)=\frac{1}{\alpha \gamma}\left((1-\frac{\gamma}{2} X^2)^{\alpha} -1\right)\,, \quad  
\partial_X \mathscr{L}=-X (1-\frac{\gamma}{2} X^2)^{\alpha-1} \,.
\end{equation}
in which $\alpha$ and $\gamma$ are two  parameters. Note that for small $\gamma$ one has the  Maxwell Lagrangian. 

Let us show that for $\gamma$  not vanishing, we may obtained  specific exact black hole solutions.
In fact, from (\ref{ln}) one gets
\begin{equation}
\label{g1}
(1-\frac{\gamma}{2} X^2)^{-2\alpha+2}=r^4 X^2\,.
\end{equation}
Again, for $\alpha=1$ one has the usual Maxwell case. Thus, we consider $\alpha \neq 1$.

The choice $\alpha=\frac{1}{2}$ leads the well known  Einstein-Born-Infeld case, studied also in \cite{GSP}. With this choice one has
\begin{equation}
\label{g2}
X^2= \frac{1}{r^4+\frac{\gamma}{2} }\,.
\end{equation}
This means that the static electric field is regular at $r=0$, Born-Infeld model. Furthermore, since

\begin{equation}
\label{g22}
\mathscr{L}=- \frac{2}{\gamma} \left( 1+r^2 X \right)\,,
\end{equation}
 the effective density reads 
\begin{equation}
\label{g33}
r^2 \rho(r)=\frac{2 r^2}{\gamma}+\frac{ 2 \sqrt{r^4+\frac{\gamma}{2}}}{\gamma}   \,.
\end{equation}

In order for this object to satisfy the WEC, it is necessary to require $\gamma > 0$, since  $\rho$ is ill defined in the  limit $\gamma \rightarrow 0$, and  no solution associated with a vanishing electromagnetic field might exist. However, since $\gamma$ is an external parameter and not an integration constant, there is no trouble in fixing it to be positive, so that Lagrangian (\ref{ln}) satisfies the WEC.

When $\gamma >0$, the  particular solution is
\begin{equation}
\label{g4}
f_1(r)=1-\frac{8\pi }{r}\int_0^r r^2_1\rho(r_1) dr_1 \,,
\end{equation}
and it may be expressed in term of Elliptic function, but it is easy to show there is no strictly  de Sitter core for $r \rightarrow 0$. In fact, condition (\ref{S1}) is violated 
\begin{equation}
\label{g4}
 \lim_{r \rightarrow 0}  r^2\rho(r)=\sqrt{2/\gamma}+\frac{2r^2}{\gamma}+O(r^4)  \,.
\end{equation}
The presence of the non vanishing constant $ \sqrt{2/\gamma} $ means that a conical singularity is present.

The other tractable case is $\alpha=\frac{3}{4}$. In fact, one has
\begin{equation}
\label{g5}
X^4r^8=(1-\frac{\gamma}{2} X^2)\,.
\end{equation}
As a consequence,
\begin{equation}
\label{g6}
X^2= \frac{1}{2r^8}\left(-\frac{\gamma}{2}+\sqrt{ \frac{\gamma^2}{4}+4 r^8}  \right)\,.
\end{equation}
For small $r$, one gets
\begin{equation}
\label{g6}
X^2\simeq  \frac{2}{\gamma}+ O(r^8)\,.
\end{equation}
Thus, the static electric field is regular in $r=0$, and the particular solution $f_1(r)$ cannot be expressed in closed form. However, a direct calculation shows that there is no de Sitter or AdS core and a conical singularity is still present.

As last example, let us consider the generalized Maxwell Lagrangian

\begin{equation}
\label{chin1}
 \mathscr{L}(X)=-\frac{1}{\xi^2}\left(\sqrt{B}-\sqrt{ X} \right)^2\,, \quad  B=\frac{A}{4\pi \xi^2} \,,
\end{equation}
with $A$ and $\xi$ given parameters. One easily gets

\begin{equation}
\label{c2}
 \sqrt{X}=\frac{\sqrt{B}r^2}{r^2+\xi^2}, \quad  
\rho=\frac{B}{r^2+\xi^2} \,.
\end{equation}

Here the WEC is satisfied as long as $B$ is positive. However, also in this case $B$ is just an external parameter; its positivity rests on the condition $A \geq 0$, but this is a safe condition, since we are allowed to impose it into the Lagrangian.

For this model, one has no conical singularity in the origin, namely a de Sitter core, and the particular solution reads
\begin{equation}
\label{g4}
f_1(r)=1-\frac{2A}{\xi^2}+ \frac{2A}{\xi r} \arctan (\frac{r}{\xi})\,.
\end{equation}
However, this particular solution  is not for large $r$ asymptotically Minkoskian.

Recall that the general solutions are $f(r)=f_1(r)-\frac{C}{r}$, and they are singular as soon as $C$ is not vanishing.

\section[toctitle]{A Reconstruction method within NED models}

In this section we present a discussion on a quite efficient reconstruction scheme. We recall that we are
working within gravity in the NED framework, namely a specific gravitational model in which the gravitational field of Einstein-Hilbert Lagrangian has as a source a suitable non linear electromagnetic field.
Within this framework, the  Einstein equation, re-written  as 
\begin{equation}
\label{ege55}
\frac{d}{dr}\left( r(f-1)\right) = - 8\pi r^2 \rho\,,
\end{equation}
 gives $\rho(r)$ once $f(r)$ is known.  The other two  two equations are

\begin{equation}
\label{r_1}
r^2 \partial_X \mathscr{L} =1 \,, \quad \mathscr{L}=\frac{X}{r^2}-\rho\,.
\end{equation}
From the above equations,  we get

\begin{equation}
\label{r_2}
X=-\frac{r^3}{2} \frac{d}{dr} \rho\,.
\end{equation}
As a consequence, one may obtain $r=r(X)$, and, making use of the second equation,   $\mathscr{L}=\mathscr{L}(X)$.

As warm up, let us start from the singular RN solution
\begin{equation}
\label{SdSRN_solution}
f(r) = 1-\frac{C}{r} + \frac{Q^2}{r^2} 
\end{equation}
Making use of (\ref{ege55}), one has
\begin{equation}
\label{HL}
\rho=\frac{Q^2}{8\pi r^4 }\,,
\end{equation}
and it is easy to show that the Maxwell Lagrangian is recovered. 

As a second less trivial example, let us consider the general solution 
\begin{equation}
\label{f_example_1}
f(r) = 1 -\frac{C}{r} -\frac{2A}{\xi} + \frac{2A}{\xi} \frac{\arctan{\left(\frac{r}{\xi}\right)}}{r} -H_0^2r^2\,.
\end{equation}
This solution is a generalization of  black hole solution obtained from a particular Horndeski Lagrangian, namely Einstein gravity with a non minimally coupled scalar field \cite{MR}. When $H_0=0$, it reduces to (\ref{g4}).

The effective density can be obtained by (\ref{ege55}), namely
\begin{equation}
\label{r-3}
\rho = \frac{A}{4\pi \xi^2(\xi^2 + r^2)}+\frac{3 H_0^2}{8\pi}\,.
\end{equation} 
Thus, from  (\ref{r_2})
\begin{equation}
\label{r-4}
X=\frac{B r^4}{(\xi^2 + r^2)^2}\,,
\end{equation} 
where 
\begin{equation}
\label{r-44}
B=\frac{A}{4\pi}>0\,.
\end{equation}
The Lagrangian can easily  be  reconstruct, and the result is

\begin{equation}
\label{chin1}
 \mathscr{L}(X)=-\frac{1}{\xi^2}\left(\sqrt{B}-\sqrt{ X} \right)^2-\frac{3H^2_0}{8\pi} \,.
\end{equation}
When $H_0=0$, one gets a result in agreement with the previous Section. Once again, the WEC is satisfied as long as we choose a suitable form for the parameters $A$ and $H_0$ (e.g. $A, \ H_0 \geq 0$).

As a last example, let us start from the following metric, $B >0$
\begin{equation}
\label{f_2}
f(r) = 1 -\frac{C}{r} +\frac{4B\pi}{r^2+\xi^2} - \frac{4B\pi}{\xi} \frac{\arctan{\left(\frac{r}{\xi}\right)}}{r}\,.
\end{equation}
With $C=0$, this is the regular black hole solution proposed by Dymnikova in \cite{Dymnikova_1}. The solution is asymptotically flat, and its regular part has de Sitter core and  no conical singularity. Let us try to reconstruct the related NED Lagrangian.
Again, from (\ref{ege55}), one has
\begin{equation}
\label{r-5}
\rho = \frac{B}{(\xi^2 + r^2)^2}\,,
\end{equation} 

and since $B$ is positive, this density clearly satisfies the WEC. In this case, also DEC is
satisfied.

Furthermore, making use of (\ref{r_2})
\begin{equation}
\label{r-6}
X = \frac{2 B r^4}{(\xi^2 + r^2)^3}\,.
\end{equation}
We may re-write it as
\begin{equation}
\label{r-7}
X(\xi^2 + r^2)^3=2B (r^2)^2\,,
\end{equation} 
and consider  $r^2$ as a function of $X$, obtained solving an algebraic equation of third order. Once we  have the solution,  the Lagrangian reads
\begin{equation}
\label{r-8}
\mathscr{L}(X) = \frac{(r^2(X) -\xi^2) X}{ 2(r^2(X))^2}\,. 
\end{equation}
However, the final expression can be written explicitly, but it is too complicate,  and it will not write down  here.  

With regard to this,  we   note   $\mathscr{L}$ may be defined by the parametric representation
\begin{equation}
\label{r-9}
\mathscr{L}(r) = \frac{(r^2 -\xi^2)}{(r^2+\xi^2)^3}\,,
\end{equation}
\begin{equation}
\label{r-10}  
X(r) = \frac{2 B r^4}{(\xi^2 + r^2)^3}\,,
\end{equation}
and this parametric representation gives
\begin{equation}
\label{r-9a}
\frac{d \mathscr{L}(r)}{d r} = \frac{Br(8\xi^2-4r^2)}{(r^2+\xi^2)^4}\,,
\end{equation}
\begin{equation}
\label{r-10a}  
\frac{d X(r)}{dr} = \frac{ B r^3(8\xi^2-4r^2)}{(\xi^2 + r^2)^4}\,.
\end{equation}
Thus, $r^2 \partial_X\mathscr{L}=1 $, namely the generalized Maxwell equation is satisfied,
and one may  evaluate the effective density $\rho=\frac{X}{r^2}-\mathscr{L} $ as a function of $r$, and  one arrives at the solution of Einstein equation   (\ref{f_2}) by a simple quadrature.

Finally we conclude this Section with the following remark. One can start with the $\mathscr{L}$ given in an implicit form
\begin{equation}
\label{r11}
X=G(\mathscr{L}) \,,
\end{equation}
where $G$ is a smooth known  function. Taking the derivative with respect to $X$, and making use of  
$r^2 \partial_X\mathscr{L}=1$, one has
\begin{equation}
\label{r12}
r^2=\partial_{\mathscr{L}}G(\mathscr{L}) \,.
\end{equation}
In principle, this gives $\mathscr{L}$ as function of $r$, and the effective density may be computed
by means
\begin{equation}
\label{r13}
r^2 \rho =G(\mathscr{L})-r^2  \mathscr{L} \,.
\end{equation}
For example, let us consider
\begin{equation}
\label{r14}
X=G(\mathscr{L})=G_0+G_1\mathscr{L}+\frac{G_2}{2} \mathscr{L}^2 \,.
\end{equation}
One has
\begin{equation}
\mathscr{L}=\frac{r^2-G_1}{G_2}\,,
\end{equation}
and
\begin{equation}
\label{r13a}
r^2 \rho =G_0-\frac{G_1^2}{2G_2}+r^2 \frac{G_1}{G_2}-\frac{r^4}{2G_2}  \,.
\end{equation}
Now in order to avoid the presence of conical singularities, one has the constraint
\begin{equation}
\label{r13b}
G_0=\frac{G_1^2}{2G_2}  \,.
\end{equation}
As a consequence, the Lagrangian is determined by the algebraic equation
\begin{equation}
\label{r14a}
X=\frac{G_1^2}{2G_2} +G_1\mathscr{L}+\frac{G_2}{2} \mathscr{L}^2 \,,
\end{equation}
and one gets
\begin{equation}
\label{r13ab}
r^2 \rho =r^2 \frac{G_1}{G_2}-\frac{r^4}{2G_2}  \,.
\end{equation}
The related general solution reads
\begin{equation}
\label{f_example_6}
f(r) = 1 -\frac{C}{r}  - 8\pi \left( \frac{r^2 G_1}{3G_2}-\frac{r^3}{10G_2}   \right) \,.
\end{equation}
To our knowledge, this is a new static spherically symmetric solution. Other solutions can be found with the same technique,  the above solution being the simplest one. Moreover, if $G_1$ and $G_2$ are negative, the WEC is satisfied. However, since $f(r)$ contains the cubic term $r^3$, for large $r$, the Kretschmann scalar will diverges like $r^2$.

We conclude this Section with some remarks on WEC. We have shown that in all the cases  studied, the parameters of the Lagrangian can be arranged such that the WEC is satisfied. Actually we trust this is a general property.

\section{Examples of non singular black holes}

In order to arrive at a non singular black hole, at least within this effective approach, a possibility consists  in assuming that the effective density depends on $f(r)$ and its derivative.  In particular, if we  assume that
\begin{equation}
 \label{ege10}
8\pi r^2\rho(r)=\frac{d}{dr}\left(\nu r^{-a} (f-1)^\gamma +Br^{b+1}  \right) \,,
\end{equation}
where $\nu$, $a$, $b$, $B$, and $\gamma$ are suitable constants, one has 
\begin{equation}
\label{ege10a}
\frac{d}{dr}(r(f-1))=-8\pi\frac{d}{dr}\left(\nu r^{-a} (f-1)^\gamma +Br^{b+1}  \right) \,.
\end{equation}
Thus
\begin{equation}
\label{ege10b}
r(f-1)+ \left(\nu r^{-a} (f-1)^\gamma +Br^{b+1}  \right)=-C \,.
\end{equation}
where $C$ is an arbitrary constant of integration. 

We start with  $\gamma=1$  This choice leads to
\begin{equation}
 \label{ege10}
f(r)=1- \frac{r^a}{\nu+r^{1+a}}\left(C+Br^{b+1} \right)  \,.
\end{equation}
In the limit $\nu \rightarrow 0$, one gets
\begin{equation}
 \label{ege11}
f(r)=1- \frac{C}{r}-Br^b \, \quad b>0  \,.
\end{equation}
When $b=2$, this reproduces the Kottler singular black hole, and $B$ is related to the cosmological constant. As a result, the parameter $\nu$ describes the regularity of the black hole

In order to deal with a non singular black hole, one has to check the existence of the de Sitter core for small $r$. One has
\begin{equation}
 \label{ege12}
f(r)=1- \frac{r^a\left(C+Br^{b+a+1} \right)}{\nu}\, \quad r \rightarrow 0  \,.
\end{equation}
Thus one has to take $a$ equal or bigger than $2$. In particular, for $a=2$ and $B=0$ one obtains the non singular Hayward black hole
\begin{equation}
 \label{ege13}
f(r)=1- \frac{C r^2}{\nu+r^3}  \,.
\end{equation}
Another interesting example is the choice $a=3$ and $b=-2$. This choice gives
\begin{equation}
 \label{ege14}
f(r)=1- \frac{r^3}{\nu+r^4}\left( C+Br^{-1} \right)  \,.
\end{equation}
This corresponds to the Lemos et al. non singular black hole \cite{BLZ}. 

Another possibility is to have $\gamma=2$, and $\nu=-\frac{1}{2\mu^2}$. In this case, one 
arrives at
\begin{equation}
 \label{ege15}
f(r)=1+ \frac{r^{1+a}}{\mu^2}\left( 1 \pm \sqrt{1+2\mu^2 r^{-2-a} (C+Br^{b+1})} \right)  \,.
\end{equation}
For the branch 
\begin{equation}
\label{ege16}
f(r)=1+\frac{r^{1+a}}{\mu^2}\left( 1 - \sqrt{1+2\mu^2 r^{-2-a}  (C+Br^{b+1})} \right)  \,.
\end{equation}
 $\mu$ is again a  regularity  parameter, since in the limit $\mu \rightarrow 0$, one has
\begin{equation}
 \label{ege17}
f(r)=1- \frac{C}{r}-Br^{b}  \,.
\end{equation}
The other branch is not a black hole solution, since $f(r)$ is always positive. 
Thus, we will consider only the solution (\ref{ege16}). 

The de Sitter core condition requires,  $a=4$ or $a>4$. As an example, we take $a=4$ and $B=0$, namely  
\begin{equation}
\label{ege18}
f(r)=1+\frac{r^{5}}{\mu^2}\left( 1 - \sqrt{1+2\mu^2 r^{-6}C} \right)  \,.
\end{equation}   
This represents a non singular black hole, which tends to Schwarzschild black hole for $\mu^2 \rightarrow 0$.
The event horizon is the positive solution of the algebraic equation given by
\begin{equation}
\label{ege20}
r_H^5=Cr^4_H -\frac{\mu^2}{2}  \,.
\end{equation}

\section{Conclusion}
In this paper, an attempt has been made to deal with so called regular black hole issue.  First we have revisited the so called NED Lagrangian approach, and we have confirmed that in general, for static spherically 
symmetric metrics, the central singularity at $r=0$ is, in general, present, unless one has physical motivations to make a suitable choice for the constant of integration.  Within this approach, we have proposed an efficient reconstruction method, and making use of it,  we have  exhibited new exact black holes solutions.   In the second part of the paper, we have followed a non Lagrangian phenomenological approach and we have discussed a  class of non singular black hole solutions.

\section{Acknowledgments}
 
We would like to thank prof. A. Ghosh, prof. L. Vanzo  for the very useful discussions, and a referee for usefull remarks.

\end{document}